\documentclass[twocolumn,prb,showpacs]{revtex4}
\usepackage{psfrag}
\usepackage{graphicx}
\usepackage{amsmath}
\usepackage{amssymb}
\usepackage{eufrak}

\begin{document}
\title{Electron transport through an interacting region: The case of a nonorthogonal\\ basis set}
\author{Kristian S. Thygesen}
\affiliation{Institut f{\"u}r Theoretische Physik, Freie
Universit{\"a}t Berlin, Arnimallee 14, D-14195 Berlin, Germany}

\date{\today}

\begin{abstract}
The formula derived by Meir and Wingreen [Phys. Rev. Lett. {\bf
68}, 2512 (1992)] for the electron current through a confined,
central region containing interactions is generalized to the case
of a nonorthogonal basis set. As the original work, the present
derivation is based on the nonequilibrium Keldysh formalism. By
replacing the basis functions of the central region by the
corresponding elements of the dual basis the lead- and central
region-subspaces become mutually orthogonal. The current formula
is then derived in this new basis, using a generalized version of
second quantization and Green's function theory to handle the
nonorthogonality within each of the regions. Finally, the
appropriate nonorthogonal form of the perturbation series for the
Green's function is established for the case of electron-electron
and electron-phonon interactions in the central region.
\end{abstract}

\pacs{73.23.-b,73.63.-b,73.63.Nm} \maketitle

\begin{section}{Introduction}\label{sec.intro}
Electron transport in nano-scale contacts is a highly active
research area. During the last decade it has become possible to
create two-terminal junctions where atomic-sized conductors are
contacted by macroscopic metal electrodes using scanning
tunnelling microscopes~\cite{tao03,rubio}, mechanically controlled
break junctions~\cite{reed97,reichert_weber02}, or
electromigration techniques~\cite{park02}. In this way
\mbox{\emph{I-V}} characteristics have been obtained for a variety
of different nano-contacts including carbon
nanotubes~\cite{nygaard}, metallic point
contacts~\cite{agrait_report}, atomic wires~\cite{yanson98}, as
well as individual molecules ranging from large organic
compounds~\cite{reichert_weber02,tao03} down to a single hydrogen
molecule~\cite{smit_nature02}.

The quantitative modelling of the electrical properties of a
nano-scale contact represents a great theoretical challenge
involving a detailed description of both the atomic- and
electronic structure of a current-carrying system out of
equilibrium. At present, the most popular approach to the problem
combines a non-equilibrium Green's function (NEGF) formalism with
\emph{ab initio} electronic structure theory. A cornerstone of
this approach is a formula giving the current through the system
in terms of the Green's function of a spatial region containing
the contact (the central region). When interactions are limited to
the central region, the current formula is an exact result,
however, in practice some approximation for the full interacting
Green's function must be invoked. For example, in the commonly
used NEGF-DFT approximation the exact Green's function is replaced
by the non-interacting Kohn-Sham Green's function defined within
Density Functional Theory~\cite{xue01,brandbyge02}.

Application of the NEGF theory to electronic transport in quantum
wires was introduced as an alternative to the
Landauer-B{\"u}ttiker formalism to treat electronic interactions.
In 1991 Hershfield and co-workers~\cite{hershfield91} derived an
expression for the current through a single interacting level (an
Anderson impurity), and the following year Meir and
Wingreen~\cite{meir_wingreen92} generalized the current formula to
the case of an arbitrary number of states in the central region
[Eq.~(6) of Ref.~\onlinecite{meir_wingreen92}]. In these studies
the system, i.e. the conductor, was partitioned into three parts
(the leads plus the central region), and the basis of the
single-particle Hilbert space was taken as an orthonormal set of
functions each belonging to one of the three regions. In practical
\emph{ab initio} calculations, however, the requirement of
localized basis functions, which is essential for the
partitioning, is difficult to combine with
orthogonality~\cite{siesta,lippert97}. It is therefore of great
practical importance to generalize the current formula to
nonorthogonal basis sets.

In this paper, a rigorous operational framework for applying
second quantization and Green's function methods in a
nonorthogonal basis is presented and used derive a generalized
current formula which is valid for nonorthogonal basis sets. The
main problem in deriving the current formula in the general case
is the lack of orthogonality between basis functions in the
central region and basis functions in the leads (due to the
assumption of localized basis functions, we can take basis
functions belonging to different leads to be non-overlapping and
thus orthogonal). We solve this problem by replacing the basis
functions of the central region by the so-called dual basis
functions which, by construction, are orthogonal to the basis
functions of the leads. The nonorthogonality of the basis within
each of the three regions is handled by applying a generalized
version of second quantization and Green's function theory.

For non-interacting electrons, the NEGF current formula takes a
particularly simple form [Eq.~(7) of
Ref.~\onlinecite{meir_wingreen92}]. This version of the formula,
which is equivalent to the well-known Landauer-B{\"u}ttiker
formula, has been applied extensively over the last years and
forms the basis of most numerical schemes addressing
phase-coherent transport in nano-scale structures. In the special
case of no interactions, Xue \emph{et al.} have derived the form
of the current formula in a nonorthogonal basis.~\cite{xue01} Xue
\emph{et al.} takes a route which is less direct than the one by
Meir and Wingreen as it involves transformations between the basis
and real space representations. In contrast, the derivation given
here follows the original work closely and is formulated entirely
within the basis set representation. More importantly, the
derivation of Xue \emph{et al.} cannot be extended to the
interacting case since it relies on a spectral representation of
the Green's function in terms of single-particle eigenstates.
Emberly and Kirczenow~\cite{emberly98} have addressed the problem
of nonorthogonality within the Landauer-B{\"u}ttiker formalism.
Here the current due to non-interacting electrons is obtained
directly from the singe-particle scattering states which are
evaluated in a Hilbert space with an energy-dependent inner
product. Finally, Fransson \emph{et al.} have studied the effects
of nonorthogonality within the tunnelling
formalism.~\cite{fransson01,fransson02} Using the Kadanoff-Baym
approach and the Hubbard operator technique, they derived an
expression for the current through a single interacting level
(Anderson model) in the weak coupling limit, taking the finite
overlap of the tunnelling wave functions into
account.~\cite{fransson02} In contrast, the derivation presented
here makes no assumptions about the type of interactions and is
not limited to the case of weak coupling between the leads and the
central region. Moreover, the use of the dual basis in the central
region simplifies the formalism significantly and avoids many
technical problems otherwise arising from the nonorthogonality
between the lead- and central region-states.

The paper is organized as follows. In Sec.~\ref{sec.nonorthogonal}
the general theory of second quantization and Green's functions in
a nonorthogonal basis is reviewed, and the concept of the dual
basis is introduced. In Sec.~\ref{sec.current} these results are
used to derive the current formula in a nonorthogonal basis, both
in the interacting and the non-interacting case.
\end{section}

\begin{section}{Nonorthogonal basis sets}\label{sec.nonorthogonal}
In this section we generalize the second quantization formalism
and elements of the Green's function theory to the case of
nonorthogonal basis sets. To avoid mathematical problems with
convergence, we assume that our single-particle Hilbert space,
$\mathcal H$, is finite dimensional. While this assumption might
seem unsatisfactory from a fundamental point of view, it will
always be true in practical applications.

\begin{subsection}{Second quantization of one- and two-body\\
  operators}\label{sec.operators}
Throughout this section let $\{\phi_i\}$ denote a -- not
necessarily orthonormal -- basis of the single-particle Hilbert
space, $\mathcal H$. The corresponding creation and annihilation
operators~\cite{definition}, $c_i^{\dagger}$ and $c_i$, acting on
the fermionic Fock space fulfill the canonical anti-commutation
relations~\cite{solovej94}
\begin{eqnarray}\label{eq.anticomm1}
\{c_i^{\dagger},c_j^{\dagger}\}&=&0\\ \label{eq.anticomm2}
\{c_i,c_j\}&=&0\\
\label{eq.anticomm3}
\{c_i,c_j^{\dagger}\}&=&S_{ij},
\end{eqnarray}
where $S_{ij}=\langle \phi_i|\phi_j\rangle$ is the overlap matrix.

To any one-body operator, $\hat A^{(1)}$, acting on $\mathcal H$,
we associate the matrix $A_{ij}=\langle \phi_i | \hat A^{(1)} |
\phi_j \rangle$. We have the following representation of $\hat
A^{(1)}$ in terms of the basis vectors $\{\phi_i\}$,
\begin{equation}\label{eq.1strepr}
\hat A^{(1)}=\sum_{ij}\mathfrak A_{ij}|\phi_i\rangle
\langle \phi_j|,
\end{equation}
where we have used Dirac's bra-ket notation and introduced the
matrix $\mathfrak A=S^{-1}AS^{-1}$. This matrix transform will
appear often throughout the text, and we shall reserve the Gothic
font for matrices that results from such a transformation. The
validity of the representation (\ref{eq.1strepr}) is easily
checked by evaluating the inner products $\langle \phi_n | \hat
A^{(1)} | \phi_m \rangle$ on both sides of the equation. The
second quantized form of $\hat A^{(1)}$, which we denote by $\hat
A$, is
\begin{equation}\label{eq.2ndrepr}
\hat A=\sum_{ij}\mathfrak A_{ij} c_i^{\dagger}c_j.
\end{equation}
The easiest way to derive this expression is to start from the well
known form of $\hat A$ in terms of some orthonormal basis,
$\{\psi_n\}$, and corresponding creation and annihilation
operators $d^{\dagger}_n,d_n$ and then expand these in terms of the
original $c^{\dagger}_i,c_i$. The explicit form of this
expansion reads
$\psi_n=\sum_{ij}(S^{-1})_{ij}\langle \phi_j|\psi_n\rangle  |\phi_i
\rangle$, which follows from Eq.~(\ref{eq.1strepr}) applied to the identity operator.

We now turn to a general two-body operator, $\hat B^{(2)}$,
defined on the two-particle Hilbert space $\mathcal
H^{(2)}=\mathcal H \otimes \mathcal H$. A basis for $\mathcal
H^{(2)}$ is provided by the tensor products $\{\phi_i \otimes
\phi_j\}$. We define the corresponding overlap matrix
$S^{(2)}_{ij,kl}=\langle \phi_i\otimes \phi_j|\phi_k \otimes
\phi_l \rangle=S_{ik}S_{jl}$, as well as the matrix
\mbox{$B_{ij,kl}=\langle \phi_i \otimes \phi_j |\hat B^{(2)} |
\phi_k \otimes \phi_l \rangle$}. As for the one-body operators we
have the first quantized representation
\begin{equation}
\hat B^{(2)}=\sum_{i,j,k,l}\mathfrak B_{ij,kl}|\phi_i \otimes \phi_j\rangle \langle \phi_k
\otimes \phi_l|,
\end{equation}
where the matrix $\mathfrak B$ is defined by $\mathfrak B=(S^{(2)})^{-1}B(S^{(2)})^{-1}$. For
later use we note that
\begin{equation}\label{eq.2bodyoverlap}
(S^{(2)})^{-1}_{ij,kl}=(S^{-1})_{ik}(S^{-1})_{jl},
\end{equation}
which can be directly verified by multiplication with $S^{(2)}$.
Finally, the following expression for the second quantized version
of $\hat B^{(2)}$ can be derived using the same technique as for
the one-body operator,
\begin{equation}\label{eq.2ndrepr2}
\hat B=\sum_{i,j,k,l}\mathfrak B_{ij,kl} c_i^{\dagger}c_j^{\dagger}c_lc_k.
\end{equation}
\end{subsection}

\begin{subsection}{Single-particle Green's functions}\label{sec.GF}
In order to fix notation we start with some well known
definitions. Given two single-particle
  orbitals $\phi_i$ and $\phi_j$ (not necessarily normalized or
  orthogonal) the retarded and advanced
  single-particle Green's functions (GFs) are defined by
\begin{eqnarray}
G^r_{ij}(t,t')&=&-i\theta(t-t')\langle \{c_i(t),c^{\dagger}_j(t')\}\rangle\\
G^a_{ij}(t,t')&=&i\theta(t'-t)\langle \{c_i(t),c^{\dagger}_j(t')\}\rangle.
\end{eqnarray}
Here the brackets $\langle \rangle$ denotes an expectation value
with respect to the equilibrium state of the system. The greater
and lesser GFs are defined by
\begin{eqnarray}\label{eq.lessergreater}
G^{<}_{ij}(t,t')&=&i\langle c^{\dagger}_j(t') c_i(t)\rangle\\
G^{>}_{ij}(t,t')&=&-i\langle c_i(t)c^{\dagger}_j(t')\rangle.
\end{eqnarray}
In fact all of these GFs can be derived from the contour-ordered
GF which is defined by
\begin{equation}
G_{ij}(\tau,\tau')=-i\langle
T_C[c_i(\tau)c^{\dagger}_j(\tau')]\rangle.
\end{equation}
Here $\tau=(t,\sigma)$ is a collection of the time variable $t$
and a branch index, $\sigma$, and $T_C$ is the contour-ordering
operator. Note that $c_i(\tau)\equiv c_i(t)$ and
$c^{\dagger}_j(\tau')\equiv c^{\dagger}_j(t')$, while the branch
indices merely serve to determine the ordering of the operators.
For more comprehensive introductions to the general GF theory the
reader is referred to
Refs.~\onlinecite{haug_jauho,bruus_flensberg,rammer_smith}.

We consider first the case where both the expectation value and
the time-evolution of the creation and annihilation operators
entering the GF is governed by a time-independent, quadratic
Hamiltonian, $\hat h$. Expressing $\hat h$ as in
Eq.~(\ref{eq.2ndrepr}), using that \mbox{$\partial_t
  c_i(t)=i[\hat h,c_i](t)$}, and Fourier transforming with respect to the time difference
$t-t'$, we obtain the equation of motion for the retarded GF
matrix in the basis $\{\phi_i\}$:
\begin{equation}\label{eq.eomgf}
(S^{-1}\omega^{+}-S^{-1}hS^{-1})G^r(\omega)=I.
\end{equation}
Here $h_{ij}=\langle \phi_i|\hat h^{(1)}|\phi_j\rangle$ where
$\hat h^{(1)}$ is the first-quantized version of $\hat h$, and
$\omega^+=(\omega+i\eta)$ with $\eta$ a small positive number
ensuring proper convergence of the Fourier integral. The same
equation holds for $G^a(\omega)$ when $\eta \to -\eta$. It is
useful to introduce another matrix quantity related to the GF and
defined by
\begin{equation}\label{eq.overlapGF}
\mathfrak G^x=S^{-1}G^xS^{-1}.
\end{equation}
As indicated by the superscript $x$ the definition applies to any
of the GFs introduced above. To have a name we shall refer to
$\mathfrak G$ as the overlap GF. Its retarded variant clearly
fulfills the following matrix equation
\begin{equation}\label{eq.matrixequation}
(S\omega^{+}-h)\mathfrak G^r(\omega)=I.
\end{equation}

Next, we ask about the form of the perturbation series of the GF
in a nonorthogonal basis. We thus consider a quantum system with a
Hamiltonian $\hat H=\hat h_0+\hat V$, where $\hat h_0$ is a simple
quadratic Hamiltonian, while $\hat V$ is a complicated one- or
two-body perturbation. In the perturbation expansion of the
contour-ordered Green's function, $G_{ij}(\tau,\tau')$, we
encounter the usual terms (generating diagrams with two external
vertices):
\begin{equation}\label{eq.term}
\int \text{d}\tau_1\cdots \text{d}\tau_n \langle T_C[\hat
c_{i,h_0}(\tau)\hat c_{j,h_0}^{\dagger}(\tau')\hat
V_{h_0}(\tau_1)\cdots \hat V_{h_0}(\tau_n)]\rangle_0.
\end{equation}
In this expression both the average and the time evolution is
governed by $\hat h_0$, i.e. $\hat X_{h_0}(\tau)=\exp(it \hat
h_0)\hat X \exp(-i t \hat h_0)$ and $\langle \hat X
\rangle_0=\text{Tr}[\hat X \exp(-\beta\hat
h_0)]/\text{Tr}[\exp(-\beta\hat h_0)]$, with $\beta=1/kT$. When
$\hat V$ is represented in terms of a nonorthogonal basis as in
Eqs.~(\ref{eq.2ndrepr}) or (\ref{eq.2ndrepr2}), the term
(\ref{eq.term}) will generate unperturbed $(n+1)$-particle GFs (or
$(2n+1)$-particle GFs if $\hat V$ is an interaction) involving
creation and annihilation operators of the nonorthogonal orbitals
$\phi_i$. As usual this $(n+1)$- or $(2n+1)$-particle GF can be
broken down into unperturbed single-particle GFs using Wick's
theorem~\cite{bruus_flensberg,haug_jauho}. Although Wick's theorem
is normally proved for orthonormal states, its validity for
nonorthogonal states can be readily verified by expanding each
creation/annihilation operator entering the $(n+1)$- or
($2n+1$)-particle GF in terms of a fixed orthonormal basis. Wick's
theorem can then be applied to each term in this expansion, which
now involves only orthonormal states, and finally the original
basis functions can be reintroduced. The perturbation series for
the matrix $G_{ij}(\tau,\tau')$ in terms of the nonorthogonal
basis, $\{\phi_i\}$, should therefore be constructed according the
usual Feynman rules, using $G^0_{ij}(\tau,\tau')$ as free
propagator and $\mathfrak V_{ij}$ (or $\mathfrak V_{ij,kl}$) as
the coupling strengths entering at the vertices. Equivalently, the
perturbation series for the overlap GF, $\mathfrak
G_{ij}(\tau,\tau')$, is obtained by evaluating each diagram using
$\mathfrak G^0_{ij}(\tau,\tau')$ as free propagator and $V_{ij}$
(or $V_{ij,kl}$) as coupling. In the case where $\hat V$ contains
interactions this follows from the identity
Eq.~(\ref{eq.2bodyoverlap}).
\end{subsection}

\begin{subsection}{Dual basis}\label{sec.dual}
Below we introduce the concept of a dual basis. The dual basis
will be used in the next section for orthogonalizing the central
region and the leads in the derivation of the current formula.
Given a general basis set, $\{\phi_i\}$, (which we refer to as the
direct basis) for the finite dimensional Hilbert space, $\mathcal
H$, there exists a dual basis, $\{\phi_{\overline{i}}\}$, with the
property
\begin{equation}
\langle \phi_i|\phi_{\overline j} \rangle =\delta_{ij}.
\end{equation}
The vectors of the dual basis can be represented explicitly in
terms of the direct basis,
\begin{equation}\label{eq.expansion}
\phi_{\overline i}=\sum_j (S^{-1})_{ji}\phi_j.
\end{equation}
We shall make the general convention that indices marked with a
bar refer to the dual basis. From the expansion
(\ref{eq.expansion}) it follows that the overlap matrix of the
dual basis is simply the inverse of $S$, i.e.
\begin{equation}\label{eq.dualoverlap}
S_{\overline i\overline j}=\langle \phi_{\overline i}|\phi_{\overline j} \rangle =(S^{-1})_{ij}.
\end{equation}
Here it is  to make a connection with the work of Fransson
\emph{et al.}, who introduce a set of creation/annihilation
operators fulfilling the anti-commutation relations
$\{c_i,c_{j}^{\dagger}\}=(S^{-1})_{ij}$ (this should be contrasted
with Eq.~(\ref{eq.anticomm3})). However, as can be seen from
Eq.~(1) of Ref.~\onlinecite{fransson02} these are simply the
creation/annihilation operators of the dual basis, and thus the
formalism of Fransson \emph{et al.} is consistent with the one
presented here.

As a final observation we note that the following relation holds
for any of the single-particle GFs
\begin{equation}\label{eq.dualbasisGF}
G^x_{\overline i \overline j}=\mathfrak G^x_{ij}.
\end{equation}
That is, the GF matrix in the dual basis equals the overlap GF in
the direct basis.
\end{subsection}

\end{section}

\begin{section}{Current formula}\label{sec.current}
  In this section we derive a formula for the electron current through a spatially
confined region possibly containing interactions.
  The derivation follows the original work of Meir and
  Wingreen~\cite{meir_wingreen92}, but is here extended to the case of
  a nonorthogonal basis set. Atomic units will be used throughout.

We consider the transport of electrons through a system which can
be divided into three regions (see Fig.~\ref{fig1}): A left lead
($L$), a right lead ($R$), and a central region ($C$). For times
$t<t_0$ the three regions are uncoupled, and the leads are in
thermal equilibrium with chemical potentials $\mu_L$ and $\mu_R$,
respectively. When the leads, which we assume to be macroscopic
yet finite in size, are coupled to the central region a current
will start to flow as the system approaches a common equilibrium.
The macroscopic size of the leads ensures that a steady state with
a constant dc-current will exist for a considerable time before
the system reaches equilibrium and the current dies out. It is the
determination of this steady state current that we address in the
following.
\begin{figure}[!h]
\includegraphics[width=0.88\linewidth]{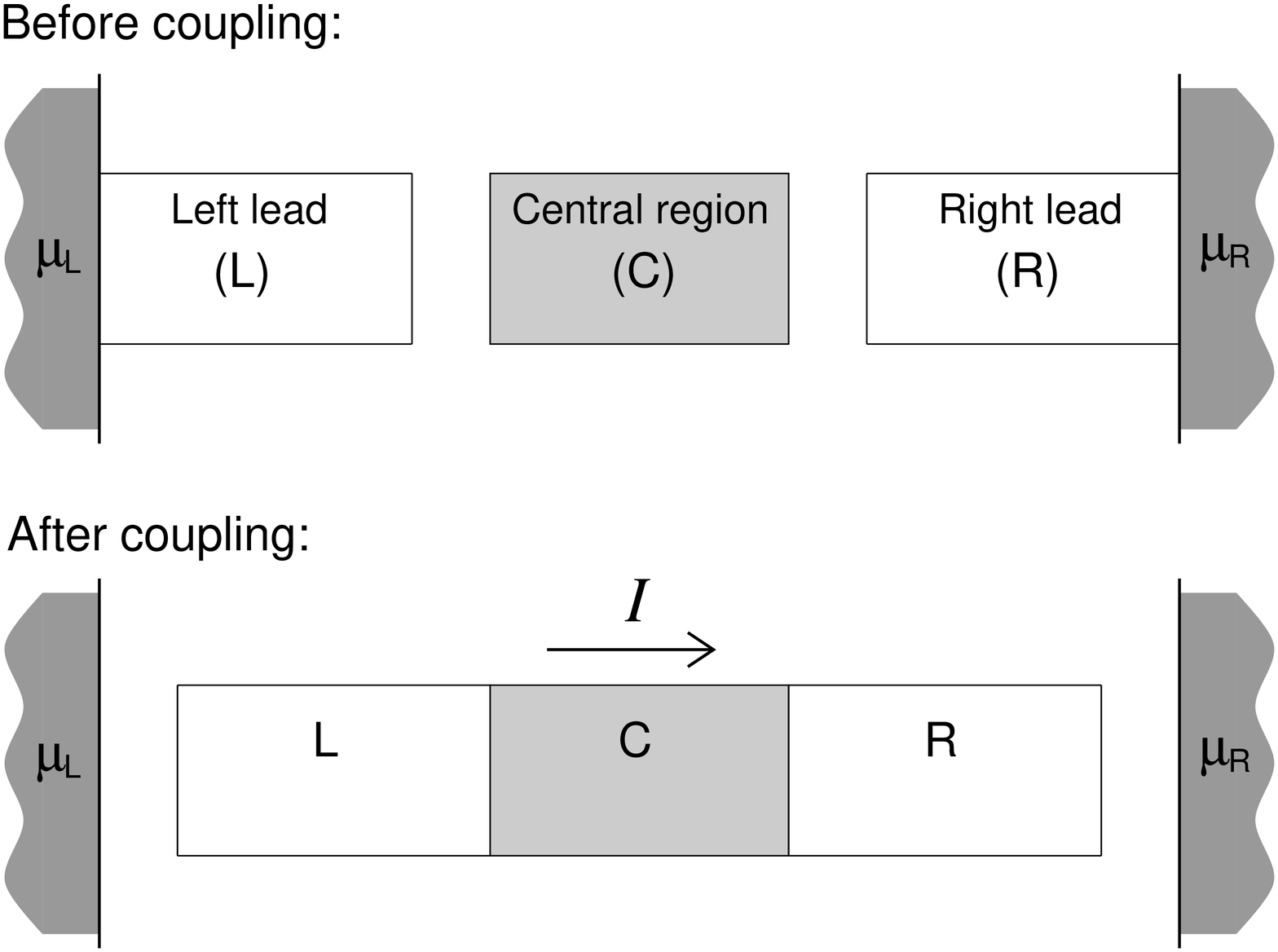}
\caption[cap.wavefct]{\label{fig1} Schematic diagram of the system
used to derive the current formula. Before the coupling between
the three regions is established, the leads are in equilibrium
with chemical potentials $\mu_L$ and $\mu_R$, respectively. Upon
coupling a current, $I$, starts to flow as the system approaches a
common equilibrium. Interactions are allowed inside the central
region ($C$).}
\end{figure}

As a basis of the single-particle Hilbert space, we take a set of
localized functions each of which can be assigned to exactly one
of the three regions, e.g. by the center position. We denote the
elements of this basis by $\phi_{\alpha i}$, where $\alpha \in
\{L,C,R\}$ specifies the region and $i$ enumerates the basis
function within the region. Since the $\phi_{\alpha i}$ are all
localized, we can safely assume that there is no overlap between
two basis functions belonging to different leads. Indeed, the
central region can always be enlarged by part of the leads until
this condition is fulfilled. We thus have $S_{Li,Rj}=0$ and
$S_{Ri,Lj}=0$ for all $i,j$, which we write compactly as
$S_{LR}=S_{RL}=0$.

The main difficulty in the derivation of the current formula is
caused by the nonorthogonality of the basis functions in the
central region and those in the leads. To overcome this problem,
we replace the basis functions in the central region by the
corresponding functions of the dual basis, i.e. $\phi_{Ci}\to
\phi_{\overline{Ci}}$. The original basis functions are maintained
in the leads. We shall refer to the resulting basis,
$\{(\phi_{Li}),(\phi_{\overline{Ci}}),(\phi_{Ri})\}$, simply as
the new basis. The introduction of the new basis amounts to a
re-definition of the subspace associated with the central region,
such that all three subspaces become orthogonal. It should be
stressed that this is merely a basis change, and thus the full
Hilbert space as well as any calculated physical quantity remains
unaffected. The nonorthogonality of basis functions within a given
region presents no serious problem, and can be handled by the
techniques of the previous section. In the following all matrix
quantities referring to the new basis will be denoted by a tilde.

The electronic Hamiltonian, $\hat H$, consists of a
non-interacting part, $\hat h$, and a part containing
interactions, $\hat V_{int}$. The physical nature of the
interactions is not important, however, we assume that $\hat
V_{int}$ only affects electrons located in the central region. In
terms of the new basis, the matrix associated with the
non-interacting part of the Hamiltonian has the generic shape:
\begin{equation}\label{eq.matrixhamiltonian}
\widetilde h=\left(
                \begin{array}{ccc}
                h_{LL} & h_{L\overline C} & 0 \\
                h_{\overline C L} & h_{\overline C \overline C} & h_{\overline C R} \\
                0 & h_{R\overline C} & h_{RR} \\
                \end{array}
               \right).
\end{equation}
Here, for example, the matrix element $\widetilde
h_{Ci,Lj}=(h_{\overline C L})_{ij}$ is given by (in the coordinate
representation),
\begin{equation}
\widetilde h_{Ci,Lj}=\int \text{d}\bold r
\phi^*_{\overline{Ci}}(\bold r)[-\frac{1}{2}\Delta_{\bold
r}+v(\bold r)]\phi_{L
  j}(\bold r),
\end{equation}
where $v(\bold r)$ is the sum of all external potentials acting on
the electron system. We shall not be concerned about the specific
form of $v(\bold r)$, as this has no importance for the general
treatment presented here. The non-interacting $\hat h$ is further
divided into the two terms, $\hat h_0$ and $\hat h_{coup}$:
\begin{equation}\label{eq.fullH}
\hat H=\hat h_0+\hat h_{coup}+\hat V_{int},
\end{equation}
where
\begin{eqnarray}\label{eq.h0}
\hat h_0&=&\hat h_0^L+\hat h_0^R+\hat h_0^C\\ &=&\sum_{\alpha\in
\{L,R\},i,j} \widetilde{\mathfrak h}_{\alpha i,
    \alpha j}c_{\alpha i}^{\dagger}c_{\alpha j}+
\sum_{i,j} \widetilde{\mathfrak
    h}_{Ci,Cj}c_{\overline{C i}}^{\dagger}c_{\overline{C j}},
\end{eqnarray}
describes each of the regions separately, and
\begin{equation}\label{eq.hcoup}
\hat h_{coup}=\sum_{\alpha\in \{L,R\},i,j} \tilde{\mathfrak
    h}_{Ci,\alpha j}c_{\overline{Ci}}^{\dagger}c_{\alpha j} +\text{H.c.},
\end{equation}
provides the coupling. As described in Sec.~\ref{sec.operators},
the matrix $\widetilde{\mathfrak h}$ is given by
$\widetilde{\mathfrak
  h}=\widetilde S^{-1}\widetilde h \widetilde S^{-1}$, where $\widetilde S$ is the
overlap matrix in the new basis,
\begin{equation}\label{eq.matrixoverlap}
\widetilde S=\left(
                \begin{array}{ccc}
                S_{LL} & 0 & 0 \\
                0 & (S^{-1})_{CC} & 0 \\
                0 & 0 & S_{RR} \\
                \end{array}
               \right).
\end{equation}
The form of $\tilde S$ within the central region follows from
Eq.~(\ref{eq.dualoverlap}). Note that in the new basis there is no
overlap between basis functions belonging to different regions of
the system, i.e. the three subspaces are orthogonal. As a
consequence, all creation/annihilation operators of a given region
commute with all creation/annihilation operators of the other
regions. This property of the new basis is crucial for what
follows.

For times $t<t_0$ the three regions are decoupled and
the state of the system is described by the density operator
\begin{equation}\label{eq.densityoperator}
\rho_0=\frac{1}{Z_0}\rho_0^L\rho_0^C \rho_0^R \otimes \rho_{ext},
\end{equation}
where $\rho_0^{\alpha}=e^{-\beta(\hat h_0^{\alpha}-\mu_{\alpha})}$
and $Z_0=\text{Tr}[\rho^0_L\rho^0_C \rho^0_R \otimes \rho_{ext}]$,
while $\rho_{ext}$ describes the state of possible external
degrees of freedom such as phonons or magnetic impurities. Notice
that the order of the density operators in
Eq.~(\ref{eq.densityoperator}) plays no role since they all
commute due to the properties of the new basis.

After the coupling has been switched on at time $t_0$, the state
$\rho_0$ will evolve according to the full Hamiltonian of
Eq.~(\ref{eq.fullH}). At the later time $t_1$ the system has
reached equilibrium and the current has died out. We assume that
there is a time interval $[t_0+\Delta t;t_1-\Delta t]$ during
which the current through the system stays constant, i.e. the
system is in a steady state. In the following we consider times
$t$ for which the system is in the steady state.

The particle current from the left region into the central region
at time $t$ is given by the time derivative of the particle number
in lead $L$~\cite{haug_jauho}
\begin{equation}\label{eq.current}
I_L(t)=\frac{\text d}{\text d t}\langle \hat{N}_L(t)
\rangle=i\langle [\hat H,\hat N_L](t)\rangle.
\end{equation}
The second quantized form of $\hat N_L$ follows from
Eqs.~(\ref{eq.1strepr}) and (\ref{eq.2ndrepr}) when $\hat A^{(1)}$
is the orthogonal projection onto the subspace spanned by the
basis functions of lead $L$,
\begin{equation}\label{eq.numberop}
\hat N_L=\sum_{ij}(S_{LL}^{-1})_{ij} c^{\dagger}_{Li}c_{Lj}.
\end{equation}
Note that only basis functions of the left lead occur in the
expression for $\hat N_L$. If we had used the original basis, the
expression would also contain creation/annihilation operators of
the regions $C$ and $R$. Since the interaction, $\hat V_{int}$, by
assumption contains creation/annihilation operators of the central
region only, the commutator in Eq.~(\ref{eq.current}) vanishes for
all terms in $\hat H$ except those coupling $L$ and $C$, i.e.
\begin{eqnarray}\label{eq.current1}
I_L&=&i\sum_{i,j}[\widetilde{\mathfrak h}_{Li,Cj}\langle
c^{\dagger}_{Li}(t)c_{\overline{Cj}}(t)\rangle-
\widetilde{\mathfrak
  h}_{Ci,Lj}\langle
c^{\dagger}_{\overline{Ci}}(t)c_{Lj}(t)\rangle]\nonumber \\ \label{eq.current2}
&=&\int \text{Tr}[\widetilde{\mathfrak
  h}_{LC}\widetilde{G}^{<}_{CL}(\omega)-\widetilde{G}^{<}_{LC}(\omega)\widetilde{\mathfrak
  h}_{CL}]\text{d}\omega
\end{eqnarray}
where the commutation relations
Eqs.~(\ref{eq.anticomm1}-\ref{eq.anticomm3}) have been used and
the lesser GF defined in Eq.~(\ref{eq.lessergreater}) has been
introduced. In the second line we have assumed that in the steady
state the GFs depend only on the time difference $t-t'$, and
moreover that the steady state exists for sufficiently long that
boundary effects associated with the switching on of the coupling
and levelling out of the current can be neglected when performing
the Fourier transform. These conditions can always be fulfilled by
increasing the size of the leads. In the following explicit
reference to the $\omega$-dependence will sometimes be omitted to
simplify the notation.

The lesser GF can be obtained from its contour-ordered counter
parts via analytic continuation as described in
Ref.~\onlinecite{haug_jauho}. Treating $\hat h_{coup}$ and $\hat
V_{int}$ perturbatively, the rules for perturbation theory in a
nonorthogonal basis (see Sec.~\ref{sec.GF}) lead to the following
Dyson equations for the lesser GF matrix
\begin{eqnarray}\label{eq.dyson11}
\widetilde{G}^{<}_{CL}&=&\widetilde{G}^{r}_{CC}\widetilde{\mathfrak
  h}_{CL}\widetilde{g}^{0,<}_{LL}+\widetilde{G}^{<}_{CC}\widetilde{\mathfrak
  h}_{CL}\widetilde{g}^{0,a}_{LL}\\ \label{eq.dyson12}
\widetilde{G}^{<}_{LC}&=&\widetilde{g}^{0,r}_{LL}\widetilde{\mathfrak
  h}_{LC}\widetilde{G}^{<}_{CC}+\widetilde{g}^{0,<}_{LL}\widetilde{\mathfrak
  h}_{LC}\widetilde{G}^{a}_{CC},
\end{eqnarray}
where $\widetilde{g}^{0}_{LL}$ is the GF of the uncoupled left
lead, i.e. the GF defined by $\hat h_0$ and $\rho_0$. Here it is
important to realize that this is \emph{not} an equilibrium GF,
since $\rho_0$ involves different chemical potentials and
therefore is not an equilibrium state. However, since the
$c_{Li},c_{Li}^{\dagger}$ commute with the corresponding operators
of region $C$ and $R$ (due to the properties of the new basis),
$\widetilde{g}^{0}_{LL}$ is in fact equal to the GF defined by
$\hat h_0^L$ and $\rho_0^L$. Since the latter describes a system
in equilibrium the fluctuation-dissipation theorem provides the
relation
\begin{equation}\label{eq.fluctdiss}
\widetilde{g}^{0,<}_{LL}(\omega)=-f_{L}(\omega)[\widetilde{g}^{0,r}_{LL}(\omega)-\widetilde{g}^{0,a}_{LL}(\omega)],
\end{equation}
where $f_L(\omega)$ is the Fermi distribution function of the left
lead. The relation (\ref{eq.fluctdiss}) introduces the equilibrium
distribution of the lead into the current formula. Since the
strict validity of this relation, as explained above, relies on
the orthogonality of the three regions, we again see the
importance of working in new basis.

For the contour-ordered GF matrix of the central region we have the Dyson
equation
\begin{equation}\label{eq.dyson2}
\widetilde{G}_{CC}=\widetilde{G}^0_{CC}+\widetilde{G}^0_{CC}[\widetilde{\Sigma}_L+\widetilde{\Sigma}_R+\widetilde{\Sigma}_{int}]\widetilde{G}_{CC},
\end{equation}
where $\widetilde{\Sigma}_{int}$ and $\widetilde{\Sigma}_{\alpha}$ are
self-energies due to the interactions and the coupling to lead
$\alpha$, respectively. Note that the former also contains
contributions from the coupling since a complete separation of the
diagrams related to the two perturbations is not possible. The
contour-ordered self-energy matrix due to the coupling to lead
$\alpha$ is given by
\begin{equation}\label{eq.selfenergy}
\widetilde{\Sigma}_{\alpha}(\omega)=\widetilde{\mathfrak
  h}_{C\alpha}\widetilde{g}^{0}_{\alpha \alpha}(\omega)\widetilde{\mathfrak
  h}_{\alpha C}.
\end{equation}
It is useful to introduce the coupling strength due to lead $\alpha$,
\begin{equation}\label{eq.gamma}
\widetilde{\Gamma}_{\alpha}(\omega)=i[\widetilde{\Sigma}^r_{\alpha}(\omega)-\widetilde{\Sigma}^a_{\alpha}(\omega)].
\end{equation}
Inserting $\widetilde G^<_{CL}$ and $\widetilde G^<_{LC}$ from
Eqs.~(\ref{eq.dyson11}),(\ref{eq.dyson12}) into the expression for
the current, Eq.~(\ref{eq.current2}), and symmetrizing,
$I=(I_L-I_R)/2$, we arrive, after some algebra, to the desired
current formula
\begin{eqnarray}\label{eq.finalcurrent}\nonumber
I&=&\frac{i}{2}\int
\text{Tr}\big[(\widetilde{\Gamma}_L-\widetilde{\Gamma}_R)\widetilde{G}^{<}_{CC}\\
&+&(f_L(\omega)\widetilde{\Gamma}_L-f_R(\omega)\widetilde{\Gamma}_R)(\widetilde{G}^{r}_{CC}-\widetilde{G}^{a}_{CC})\big
]\text{d}\omega.
\end{eqnarray}
Eq.~(\ref{eq.finalcurrent}) is formally equivalent to the
corresponding formula valid in an orthonormal
basis~\cite{meir_wingreen92}, and indeed it reduces to it when the
basis is orthonormal. However, it should be remembered that all
quantities entering Eq.~(\ref{eq.finalcurrent}) refer to the new
basis (as indicated by the tildes), and that the coupling
strengths, $\Gamma_L$ and $\Gamma_R$, involve
$\widetilde{\mathfrak h}$ instead of $\widetilde h$.

As the dual basis in most cases is not explicitly known it is
desirable to re-express Eq.~(\ref{eq.finalcurrent}) in terms of
the original basis, that is, in terms of untilted quantities. For
the GFs of the central region we have the simple relation
\begin{equation}\label{eq.firstequivalence}
\widetilde{G}^x_{CC}(\omega)\equiv G^x_{\overline C \overline
C}(\omega)=\mathfrak{G}^x_{CC}(\omega),
\end{equation}
where the second equality follows directly from
Eq.~(\ref{eq.dualbasisGF}). As for the $\Gamma$s we note that this
relation holds in particular when $\hat V_{int}=0$. In this case
Eq.~(\ref{eq.matrixequation}) establishes that
$\mathfrak{G}^r_{CC}=[(\omega^+ S-h)^{-1}]_{CC}$. On the other
hand $\widetilde{G}^r_{CC}$ can be obtained from the Dyson
equation (\ref{eq.dyson2}) using that $\widetilde
G^{0,r}_{CC}=[\omega^+
\widetilde{S}_{CC}^{-1}-\widetilde{S}_{CC}^{-1}\widetilde{h}_{CC}\widetilde{S}_{CC}^{-1}]^{-1}$
which in turn follows from Eq.~(\ref{eq.eomgf}). In terms of
self-energies we thus have
\begin{eqnarray}\label{eq.expr1}
\mathfrak{G}^r_{CC}&=&\big [\omega^+
S_{CC}-h_{CC}-\Sigma^r_{L}-\Sigma^r_{R}\big ]^{-1}\\ \label{eq.expr2}
\widetilde{G}^r_{CC}&=&\big [\omega^+
\widetilde{S}_{CC}^{-1}-\widetilde{S}_{CC}^{-1}\widetilde{h}_{CC}\widetilde{S}_{CC}^{-1}-\widetilde{\Sigma}^r_{L}-\widetilde{\Sigma}^r_{R}\big
]^{-1}
\end{eqnarray}
where~\cite{datta_book}
\begin{equation}
\Sigma^r_{\alpha}=(\omega^+ S_{C \alpha}-h_{C\alpha})[\omega^+
S_{\alpha \alpha}-h_{\alpha \alpha}]^{-1}(\omega^+ S_{\alpha C}-h_{\alpha C}).
\end{equation}
By equating Eqs.~(\ref{eq.expr1}) and (\ref{eq.expr2}), it follows
after some matrix algebra, that
\begin{eqnarray}\label{eq.secondequivalence}\nonumber
\widetilde{\Gamma}_{\alpha}&=&i[\Sigma^r_{\alpha}-\Sigma^a_{\alpha}]-2i\eta
S_{C\alpha}S_{\alpha \alpha}^{-1}S_{\alpha C} \\ &\to&
\Gamma_{\alpha}\text{ as }\eta \to 0,
\end{eqnarray}
where $\Gamma_{\alpha}=i[\Sigma_{\alpha}^r-\Sigma_{\alpha}^a]$ and
$i\eta$ is the imaginary part of $\omega^+$.
Eq.~(\ref{eq.secondequivalence}) holds, of course, also when
interactions are present.

With equations (\ref{eq.firstequivalence}) and
(\ref{eq.secondequivalence}) we can state our main result, namely
an expression for the current in terms of quantities which are all
evaluated in terms of the original nonorthogonal basis:
\begin{eqnarray}\label{eq.finalcurrent2}\nonumber
I&=&\frac{i}{2}\int
\text{Tr}\big[(\Gamma_L-\Gamma_R)\mathfrak{G}^{<}_{CC}\\
&+&(f_L(\omega)\Gamma_L-f_R(\omega)\Gamma_R)(\mathfrak{G}^{r}_{CC}-\mathfrak{G}^{a}_{CC})\big
]\text{d}\omega.
\end{eqnarray}

\begin{subsection}{Non-interacting electrons}
In the special case of non-interacting electrons the Keldysh
equation~\cite{haug_jauho} for the lesser GF of the central region
in combination with
Eqs.~(\ref{eq.fluctdiss}),(\ref{eq.selfenergy}), and
(\ref{eq.gamma}) yields
\begin{eqnarray}
\widetilde{G}_{CC}^<&=&\widetilde{G}_{CC}^r(\widetilde{\Sigma}^<_{L}+\widetilde{\Sigma}^<_{R})\widetilde{G}_{CC}^a\\
&=&\widetilde{G}_{CC}^r[f_L(\omega)\widetilde{\Gamma}_L+f_R(\omega)\widetilde{\Gamma}_R]\widetilde{G}_{CC}^a.
\end{eqnarray}
(The Keldysh equation relies only on the temporal properties of
the various GFs and is therefore valid in any single-particle
basis). Substituting this relation into
Eq.~(\ref{eq.finalcurrent}) and using the general result
$G^r-G^a=G^{>}-G^{<}$, we obtain the following Landauer-type
formula for the current,
\begin{equation}\label{eq.nonintcurrent}
I=\int  [f_L(\omega)-f_R(\omega)]\text{Tr}\big [\widetilde{G}_{CC}^r
\widetilde{\Gamma}_L\widetilde{G}_{CC}^a
\widetilde{\Gamma}_R\big ]\text{d}\omega.
\end{equation}
By virtue of the identities (\ref{eq.firstequivalence}) and
(\ref{eq.secondequivalence}) we can again re-express
Eq.~(\ref{eq.nonintcurrent}) in terms of quantities of the
original basis:
\begin{equation}
I=\int  [f_L(\omega)-f_R(\omega)]\text{Tr}\big [\mathfrak{G}_{CC}^r
\Gamma_L\mathfrak{G}_{CC}^a\Gamma_R\big ]\text{d}\omega.
\end{equation}
This is the celebrated "trace-formula" which has been widely used
for numerical calculations of coherent transport. The same formula
has previously been derived by Xue \emph{et al.} using a somewhat
different approach involving transformations to a real space
representation.
\end{subsection}

\begin{subsection}{Interactions}\label{sec.interactions}
In the presence of interactions in the central region, the current
formulas (\ref{eq.finalcurrent}),(\ref{eq.finalcurrent2}) are
exact provided the full interacting Green's function, $\widetilde
G_{CC}$ (=$\mathfrak{G}_{CC}$), is known. In this section we
address the evaluation of the interacting GF within perturbation
theory given a nonorthogonal basis. To avoid any confusion we
denote the GFs evaluated in the presence of the coupling to the
leads, but without the interaction, by the superscript "$ni$". In
the following we use the idea of the "new basis" introduced in the
beginning of Sec.~\ref{sec.current} without further explanations.

\begin{subsubsection}{Electron-electron interactions}
Assume that the electrons located in
  the central region can interact through a two-body
  potential, $\hat V^{(2)}$. The direct basis of $\mathcal H^{(2)}$ consists of the tensor
  products $\{\phi_{\alpha i}\otimes \phi_{\beta j}\}$, where
  $\alpha,\beta\in\{L,C,R\}$, while the new basis is obtained by
  replacing $\phi_{Ci}$ by its dual $\phi_{\overline{Ci}}$.
The limitation of interactions to the central region means
mathematically that the matrix element $V_{\alpha i \beta j,\gamma
k \delta
  l}=\langle \phi_{\alpha i}\otimes
  \phi_{\beta j}|\hat V^{(2)}|\phi_{\gamma k}\otimes \phi_{\delta l}\rangle$
  is non-zero only when all $\alpha,\beta,\gamma,\delta=S$.
  Since the single-particle basis functions used in the lead regions are the
  same in the new and the original basis, the same holds for the matrix
  $\widetilde{V}$, as well as for the matrix
  $\widetilde{\mathfrak
    V}=(\widetilde{S}^{(2)})^{-1}\widetilde{V}(\widetilde{S}^{(2)})^{-1}$. The first claim can be verified
    by expanding $\phi_{\overline{Ci}}$ in terms of the direct basis $\{(\phi_{Li}),(\phi_{Ci}),(\phi_{Ri})\}$
    according to Eq.~(\ref{eq.expansion}), and the second then follows from direct calculation using
  Eqs.~(\ref{eq.2bodyoverlap}) and (\ref{eq.matrixoverlap}).

As described in Sec.~\ref{sec.operators}, the second quantized
version of $\hat V^{(2)}$ in the new basis reads:
\begin{equation}
\hat V_{int}=\sum_{ijkl}\widetilde{\mathfrak
  V}_{CiCj,CkCl}c^{\dagger}_{\overline{Ci}}c^{\dagger}_{\overline{Cj}}c_{\overline{Cl}}c_{\overline{Ck}}.
\end{equation}
By direct calculation it can be shown that $\widetilde{\mathfrak
  V}=V$, and thus
\begin{equation}
\hat
V_{int}=\sum_{ijkl}V_{CiCj,CkCl}c^{\dagger}_{\overline{Ci}}c^{\dagger}_{\overline{Cj}}c_{\overline{Cl}}c_{\overline{Ck}}.
\end{equation}
Treating $\hat V_{int}$ as a perturbation to the non-interacting
Hamiltonian, $\hat h=\hat h_0 +\hat h_{coup}$, the corrections
(diagrams) to the non-interacting GF of the central region should
thus be evaluated according to the
  usual Feynman rules using $V$ as the coupling matrix and
  $\widetilde{G}^{ni}_{CC}$ as the free propagator, see last paragraph of Sec.~\ref{sec.GF}.
  From Eq.~(\ref{eq.firstequivalence}) we get that
$\widetilde{G}^{ni}_{CC}=\mathfrak
  G^{ni}_{CC}$, and consequently we have established the
  perturbation series for the central
  object of the current formula, $\widetilde{G}_{CC}$, in terms
  of $V$ and $\mathfrak
  G^{ni}_{CC}$, i.e. quantities referring to the original basis.
\end{subsubsection}

\begin{subsubsection}{Electron-phonon interactions}
The interaction between the electron system and a single
vibrational mode is, to first order in the ion displacements,
described by an operator of the generic form:
\begin{equation}\label{eq.el-ph}
\hat V_{int}=v^{(1)}(\bold r)[b+b^{\dagger}].
\end{equation}
Here $b^{\dagger},b$ are creation and annihilation operators
acting on the phonon system, and $v^{(1)}(\bold r)$ is a one-body
potential acting on the electrons. $v^{(1)}$ is obtained by
differentiating the electron-ion potential, $V_{el-ion}(\bold
r,\bold R)$, with respect to the ion-coordinates in the direction
of the vibration under consideration, see e.g.
Ref.~\onlinecite{bruus_flensberg}. The restriction of interactions
to the central region implies that $v_{\alpha i,\beta j}=\langle
\phi_{\alpha i}|v^{(1)}|\phi_{\beta j}\rangle$ is non-zero only
when $\alpha,\beta=C$. Physically this means that the vibration
does not distort the potential felt by an electron outside the
central region. According to Eq.~(\ref{eq.2ndrepr}), the matrix
that is relevant for the second quantized form of $v^{(1)}$ in the
new basis, is $\widetilde{\mathfrak
v}=\widetilde{S}^{-1}\widetilde{v} \widetilde{S}^{-1}$. Using the
expansion (\ref{eq.expansion}) and the block diagonal form of
$\widetilde S$ (see Eq.~(\ref{eq.matrixoverlap})), it is
straightforward to establish that $\widetilde{\mathfrak v}=v$. The
second quantized version of Eq.~(\ref{eq.el-ph}) thus reads
\begin{equation}\label{eq.el-ph2}
\hat
V_{int}=\sum_{ij}v_{Ci,Cj}c^{\dagger}_{\overline{Ci}}c_{\overline{Cj}}[b+b^{\dagger}].
\end{equation}
Considering $\hat V_{int}$ as a perturbation to the
non-interacting Hamiltonian, $\hat h=\hat h_0 +\hat h_{coup}$, the
diagrams generated by the interaction should thus be evaluated
using $v$ as the coupling matrix and
  $\widetilde{G}^{ni}_{CC}$ ($=\mathfrak G^{ni}_{CC}$) as the free propagator, see last paragraph of Sec.~\ref{sec.GF}.
\end{subsubsection}
\end{subsection}
\end{section}

\begin{section}{Summary}
The use of localized basis sets in electronic structure
calculations calls for general formulations of applied physical
theories, taking the nonorthogonality of the basis into account.
In this paper, I have presented a general form of the second
quantization formalism and Green's function theory which is valid
in a nonorthogonal basis, and used it to obtain a nonorthogonal
version of Eqs.~(6) and (7) in Ref.~\onlinecite{meir_wingreen92}
for the current through an interacting electron region. The main
problem in deriving the generalized current formula, namely that
the lead subspaces are not orthogonal to the central region
subspace, was solved by replacing the basis functions of the
central region by the corresponding elements of the dual basis.
This simply amounts to a basis change, under which the central
region subspace becomes orthogonal to the leads. The
nonorthogonality of the basis functions within each of the regions
was handled by applying the generalized form of the second
quantization and Green's function formalisms. Finally, the
appropriate nonorthogonal form of the perturbation expansion for
the Green's function was established for the case of
electron-electron and electron-phonon interactions in the central
region.
\end{section}

%%%%%%% References

\bibliographystyle{apsrev}

\begin{thebibliography}{33}
\expandafter\ifx\csname natexlab\endcsname\relax\def\natexlab#1{#1}\fi
\expandafter\ifx\csname bibnamefont\endcsname\relax
  \def\bibnamefont#1{#1}\fi
\expandafter\ifx\csname bibfnamefont\endcsname\relax
  \def\bibfnamefont#1{#1}\fi
\expandafter\ifx\csname citenamefont\endcsname\relax
  \def\citenamefont#1{#1}\fi
\expandafter\ifx\csname url\endcsname\relax
  \def\url#1{\texttt{#1}}\fi
\expandafter\ifx\csname urlprefix\endcsname\relax\def\urlprefix{URL }\fi
\providecommand{\bibinfo}[2]{#2}
\providecommand{\eprint}[2][]{\url{#2}}


\bibitem{tao03}
        B.~Xu and N.~J. Tao
        Science {\bf 301}, 1221 (2003)

\bibitem{rubio}
        G. Rubio, N. Agrait, and S. Viera,
        Phys. Rev. Lett. {\bf 76}, 2302 (1996)

\bibitem{reed97}
        M.~A. Reed, C.~Zhou, C.~J. Muller, T.~P. Burgin and J.~M. Tour
        Science {\bf 278}, 252 (1997)

\bibitem{reichert_weber02}
        J.~Reichert, R. Ochs, D. Beckman, H.~B. Weber, M. Mayor and
        H.~v. L{\"o}hneysen
        Phys. Rev. Lett. {\bf 88}, 176804 (2002)

\bibitem{park02}
        J. Park \emph{et al.},
        Nature {\bf 417}, 722 (2002)

\bibitem{nygaard}
        J. Nyg{\aa}rd, D. H. Cobden, and P. E. Lindelof,
        Nature {\bf 408}, 342 (2000)

\bibitem{agrait_report}
        N.~Agrait, A.~L. Yeyati and J.~M. van Ruitenbeek,
        Physics Reports, {\bf 377}, 81-279 (2003)

\bibitem{yanson98}
  A.~I.~Yanson, G.~Rubio~Bollinger, H.~E.~van der Brom, N.~Agrait, and J.~M.~Ruitenbeek
  Nature (London) {\bf 395}, 783 (1998).

\bibitem{smit_nature02}
        R.~H.~M.~Smit, Y.~Noat, C.~Untiedt, N.~D.~Lang, M.~C.~van
        Hemert and J.~M.~van Ruitenbeek,
        Nature {\bf 419}, 906 (2002).

\bibitem{xue01}
        Y.~Xue, S.~Datta and M.~A. Ratner,
        Chem. Phys. {\bf 281}, 151 (2001).

\bibitem{brandbyge02}
        M. Brandbyge, J.~L. Mozos, P. Ordej{\'o}n, J. Taylor and K.
        Stokbro,
        Phys. Rev. B {\bf 65}, 165401 (2002).

\bibitem{hershfield91}
        S. Hershfield, J. H. Davies, and J. W. Wilkins,
        Phys. Rev. Lett. {\bf 67}, 3720 (1991).

\bibitem{meir_wingreen92}
        Y. Meir and N. S. Wingreen,
        Phys. Rev. Lett. {\bf 68}, 2512 (1992).

\bibitem{siesta}
        J. M. Soler, E. Artacho, J. D. Gale, A. Garcia, J.
        Junquera, P. Ordej{\'o}n, and D. S{\'a}nchez-Portal,
        Journal of Physics: Cond. Mat. {\bf 14}, 2745 (2002)

\bibitem{lippert97}
        R. A. Lippert, T. A. Arias, and A. Edelman,
        J. Comp. Phys. {\bf 140}, 278 (1997)

\bibitem{emberly98}
        E. Emberly and G. Kirczenow,
        Phys. Rev. Lett. {\bf 81}, 5205 (1998)

\bibitem{fransson01}
        J. Fransson, O. Eriksson, and I. Sandalov, Phys. Rev. B {\bf 64}, 153403 (2001).

\bibitem{fransson02}
        J. Fransson, O. Eriksson, and I. Sandalov, Phys. Rev. B {\bf 66}, 195319 (2002).

\bibitem{definition}
        \emph{Definition of creation and annihilation operators.} Given any orbital, $f \in \mathcal H$,
        we first define the operators $C^{\dagger}(f)$ and $C(f)$ on the pure (unsymmetrized) tensor products
        by
        $C^{\dagger}(f)f_1 \otimes \ldots \otimes f_N=\sqrt{N+1}f\otimes f_1 \otimes \ldots \otimes f_N$ and $C(f)f_1 \otimes \ldots \otimes f_N=\sqrt{N}\langle f | f_1\rangle f_2 \otimes \ldots \otimes f_N$,
        respectively. $C^{\dagger}(f)$ and $C(f)$ are extended by
        linearity to the Fock space, $\mathcal F=\bigoplus_{n=1}\mathcal H^{(n)}$.
        The operators $c^{\dagger}_i$ and $c_i$ are defined
        on the anti-symmetric Fock space, $\mathcal F=P_- \mathcal F$, by $c^{\dagger}_i=P_-
        C^{\dagger}(\phi_i)$ and $c_i=P_-C(\phi_i)$, where $P_-$ is the anti-symmetrizing projector.

\bibitem{solovej94}
        V. Bach, E. H. Lieb, and J. P. Solovej,
        J. Stat. Phys. {\bf 76}, 3 (1994).


\bibitem{haug_jauho}
        H. Haug and A. -P. Jauho,
        \emph{Quantum Kinetics in Transport and Optics of
        Semiconductors}, Springer (1998).

\bibitem{bruus_flensberg}
        H. Bruus and K. Flensberg,
        \emph{Many-body quantum theory in condensed matter physics - an introduction}, Oxford University Press (2004).

\bibitem{rammer_smith}
        J. Rammer and H. Smith,
        Rev. Mod. Phys. {\bf 58}, 323 (1986).


\bibitem{datta_book}
        S.~Datta,\emph{Electronic Transport in Mesoscopic Systems}
        Cambridge (1995).




\end{thebibliography}

\end{document}